\documentclass[twocolumn,nofootinbib,aps,prc,amsfonts,pdftex,floatfix,superscriptaddress]{revtex4-1}
\usepackage{graphicx}  
\usepackage{amsmath, amssymb, bm}   % AMS packages  
\usepackage[normalem]{ulem}
\usepackage[dvips]{color}
\usepackage{natbib}
\usepackage{xcolor}
\usepackage{hyperref}
\hypersetup{
    colorlinks,
    linkcolor={red!50!black},
    citecolor={blue!50!black},
    urlcolor={blue!80!black}
}
\usepackage{slashed}
\def\){\right)} 
\def\({\left(} 
\def\]{\right]} 
\def\[{\left[}

\begin{document}

\title{Effective-field-theory predictions of the muon-deuteron capture rate}

\author{%
Bijaya Acharya}
\email{acharya@uni-mainz.de}
\affiliation{Institut f\"{u}r Kernphysik and PRISMA Cluster of Excellence,
  Johannes Gutenberg-Universit\"{a}t Mainz, 55128 Mainz, Germany}
\affiliation{Department of Physics and Astronomy, University of
  Tennessee, Knoxville, TN 37996, USA}

\author{%
Andreas Ekstr\"{o}m}
\email{andreas.ekstrom@chalmers.se}
\affiliation{Department of Physics, Chalmers University of Technology,
  SE-412 96 G\"{o}teborg, Sweden}

\author{%
Lucas Platter}
\email{lplatter@utk.edu}
\affiliation{Department of Physics and Astronomy, University of
  Tennessee, Knoxville, TN 37996, USA}
\affiliation{Physics Division, Oak Ridge National Laboratory, Oak
  Ridge, TN 37831, USA}

% * Abstract
\begin{abstract}

  We quantify the theoretical uncertainties of chiral
  effective-field-theory predictions of the muon-deuteron capture
  rate. Theoretical error estimates of this low-energy process are 
  important for a reliable interpretation of forthcoming experimental
  results by the MuSun collaboration. Specifically, we estimate the
  three dominant sources of uncertainties that impact theoretical
  calculations of this rate: those resulting from uncertainties in the
  pool of fit data used to constrain the coupling constants in the
  nuclear interaction, those due to the truncation of the effective
  field theory, and those due to uncertainties in the axial radius of
  the nucleon. For the capture rate into the ${}^1S_0$ channel, we
  find an uncertainty of approximately $4.6~s^{-1}$ due to the
  truncation in the effective field theory and an uncertainty of
  $3.9~s^{-1}$ due to the uncertainty in the axial radius of the
  nucleon, both of which are similar in size to the targeted
  experimental precision of a few percent.

\end{abstract}
 
\date{\today}
\maketitle

% * Introduction
\section{\bf Introduction}
\label{sec:intro}
Effective field theories (EFTs) have become a widely used tool in
particle and nuclear physics. They are used to obtain a systematic
low-energy expansion of observables when a separation of scales is
present in a given problem.  In particular, chiral EFT had a
transformative effect on low-energy nuclear theory
\cite{Bedaque:2002mn,Epelbaum:2008ga, Hammer:2012id} by providing a
clear path towards a nuclear Hamiltonian that can describe the
properties of atomic nuclei to high accuracy. Within this framework,
nucleons and pions are the degrees of freedom used to build up the
nuclear potential that is used to describe the spectra of nuclei. The
expansion parameter $Q$ of chiral EFT is given by ${\rm
  max}(m_\pi/\Lambda_b, q/\Lambda_b)$ where $m_\pi$ denotes the pion
mass, $q$ a low momentum scale and $\Lambda_b$ denotes the breakdown
scale of the theory, which is expected to be comparable to the
lightest degree of freedom not taken into account in the theory. An
additional advantage over previous approaches to the internuclear
potential is that EFT also provides clear guidance on how to construct
the coupling to external sources. Indeed, the electroweak current is
also calculated order-by-order in a low-energy expansion in chiral EFT
and thus shares a large number of low-energy constants (LECs) with the
nuclear potential. Thus, chiral dynamics constrains the form of the
nuclear currents significantly.

Uncertainty quantification of theoretical calculations is particularly
important in the nuclear electroweak sector where observables that are
very challenging, or even impossible, to measure experimentally serve
as input to astrophysical models. Fortunately, uncertainty
quantification was one of the initial promises of EFT
calculations. However,
it should be pointed out that there remain several open questions on
the meaning and understanding of renormalization group invariance of
chiral EFT~\cite{PhysRevLett.114.082502,Epelbaum2013, Nogga:2005hy}
and therefore also the interpretation of truncation errors.

In this paper we build on recent progress in uncertainty
quantification for EFTs~\cite{Carlsson:2015vda, Hoferichter:2015hva,
  Furnstahl:2015rha, PhysRevC.91.054002, 0954-3899-42-3-034003} and
present new results for the different sources of theoretical
uncertainties in the EFT description of muon capture on the deuteron,
{\it i.e.} the process
\begin{equation}
\label{eq:reaction}
\mu^- + d \rightarrow \nu_\mu + n + n~.
\end{equation}
Currently, the MuSun Collaboration is performing an experiment at the
Paul Scherrer Institut to measure the rate of this reaction to
percentage precision~\cite{Andreev:2010wd}. This will be the first
precise measurement of a weak nuclear process in the two-nucleon
($NN$) system, and the aim is to
  determine the LEC $c_D$ that parameterizes the strength of the
  short-distance part of the axial two-body current as well as the
  one-pion-exchange contact-term in the leading three-nucleon ($NNN$)
  interaction in EFT approaches to nuclear forces and currents.

Muon capture on the deuteron has long been expected to
provide understanding of the electroweak nuclear operator 
(see Ref. \cite{Measday:2001yr} and references therein).
A first chiral EFT calculation of muon capture into the
neutron-neutron ($nn$) singlet $S$-wave was carried out by Ando {\it et
  al. }\cite{Ando:2001es}. More recently, more complete calculations of
the muon capture rate were carried out in Refs.~\cite{Marcucci:2011jm, 
Marcucci:2010ts,Golak:2016zcw}.

Here, we focus on the three dominant sources of uncertainties of an
EFT calculation of the capture rate: those resulting from
uncertainties in the the nucleon-nucleon scattering database, those
due to the truncation of the EFT and those due to uncertainties in the
nucleon axial form factor. We will focus on capture from the $S$-wave
doublet state of the muonic deuterium atom to the singlet $S$-wave
state of the $nn$ system, $\Gamma_D^{{}^1S_0}$, which is the only
contribution to $\Gamma_D$ that is relevant to the contact part of the
axial current. Furthermore, this part can be extracted by
  subtracting from $\Gamma_D$ the higher partial wave contributions
  calculated in
  Refs.~\cite{Marcucci:2011jm,Marcucci:2010ts,Golak:2016zcw}. While
  these contributions have theoretical uncertainties of their own,
  they are not sensitive to physics at range shorter than that of pion
  exchange at the chiral order we are operating at.

We follow two approaches. (i) We use a family of 42 potentials at
order $Q^3$ that have been fitted at 7 different regulator cutoffs
$\Lambda$ in the range $450-600$ MeV to 6 different $T_{\rm lab}$
ranges in the $NN$ scattering data base. The LECs in this family of
$NN$+$NNN$ interactions were simultaneously fitted to pion-nucleon
($\pi N$) and selected $NN$ scattering data, the energies and charge
radii of $^{2,3}$H and $^{3}$He, the quadrupole moment of $^{2}$H, as
well as the comparative $\beta$-decay half-life of $^{3}$H. A simple
momentum-dependent error term with EFT-like scaling was included in
the fits to scattering data, and all 42 potentials reproduce the pool
of fit data equally well, see Ref. ~\cite{Carlsson:2015vda} for
details. Clearly, calculating the muon-capture rate with this family
of interactions probes an important component of the total theoretical
uncertainty. (ii) We also use a set of chiral interactions with
regulator cutoff $\Lambda=500$ MeV at orders $Q^{0},Q^{2},Q^{3}$ with
the sub-leading $\pi N$ couplings $c_1,c_3,c_4$ according to the
precise Roy-Steiner analysis presented in
Refs.~\cite{Hoferichter:2015hva, Siemens:2017jb}. The corresponding
$NN$ contact-potentials of this set of interactions are constrained to
reproduce the $NN$ phase shifts of the Granada
PWA~\cite{PhysRevC.88.064002} up to 200 MeV lab scattering energy as
well as the binding energy and radius of the deuteron. This second
class of interactions enables us to parameterize $\Gamma_D^{{}^1S_0}$
in terms of only one LEC, either $\hat{d}_{R}$ or $c_D$, which are
related by $\hat d_R = -\textstyle{\frac{m }{4g_A\Lambda_b}}c_D
+\frac{1}{3}\hat c_3 + \frac{2}{3}\hat c_4 +\frac{1}{6}$, where $\hat
c_i \equiv c_i m$ and $m$ is the nucleon
mass~\cite{Marcucci:2011jm,0000000002995761,Baroni:2015uza,Krebs:2016rqz,Schiavilla:cDvsdR}
~\footnote{An error in the coefficient of $c_D$ in
    Ref.~\cite{0000000002995761} was recently corrected by
    Ref.~\cite{Schiavilla:cDvsdR}. The LECs of
  Ref.~\cite{Carlsson:2015vda} have been re-optimized with a corrected
  relation between the LECs $c_D$ and $\hat
  d_R$~\cite{Schiavilla:cDvsdR}.  The new
  values~\cite{Ekstrom:tobepublished} are used throughout in this
  work.}.  Indeed, after extracting the LEC $c_E$ of the leading
$NNN$ contact from the energy and radius of $^{3}$H and $^3$He, the
three-nucleon force is completely predicted up to order $Q^3$ by
$\Gamma_D^{{}^1S_0}$.

In the following we show that the capture rates extracted from
approaches (i) and (ii) agree with each other. Furthermore, we discuss
the relative size of the uncertainties of our predictions that arise
from the afore-mentioned sources, and their implications for the
interpretation of the impending experimental MuSun results.

% * 1S_0 capture rate
\section{\bf The $^1S_0$ capture rate}
\label{sec:capture_rate}
At nuclear energies, the charge-changing weak interaction Hamiltonian
$\hat H_W$ can be written in terms of the leptonic and the nuclear
weak current operators as
\begin{equation}
\label{eq:weak_current_operator}
 \hat H_W = \frac{G_V}{\sqrt{2}}\int\mathrm{d}^3x \left[j_\alpha(\mathbf{x}) J^\alpha (\mathbf{x})+\mbox{h.c.}\right],
\end{equation}
where $G_V$ is the vector coupling constant which is related to 
the Fermi coupling constant $G_F$ and the Cabibbo mixing angle $\theta_C$ by $G_V=G_F\cos\theta_C$, and 
``h.c.'' stands for the Hermitian conjugate of the preceding term.
The matrix element of the leptonic weak current operator $j^\alpha$
%between the leptonic states
is
$l^\alpha\,e^{-i \mathbf{q}\cdot\mathbf{x}}$, where $l^\alpha$ is the
Dirac current of the leptons.  The matrix element for the
process in Eq.~\eqref{eq:reaction} can then be written as
\begin{align}
\label{eq:tfi}
 T_{fi} = \frac{G_V}{\sqrt{2}}\,\phi_{1S}(\bm{0}) \sum_{s_\mu s_d} 
 & \langle \frac{1}{2}s_\mu,\,1 s_d \vert (\frac{1}{2}1)\frac{1}{2}\,s_{\mu d} \rangle\,l^\alpha(h,s_\mu) \nonumber\\
 & \qquad \langle\psi_{nn}\vert J_\alpha(\mathbf{q}) \vert \psi_d; s_d \rangle\,,
\end{align}
where
$\phi_{1S}(\bm{0})=[\alpha m_\mu m_d/(m_\mu+m_d)]^{3/2}/\pi^{1/2}$ is
the ground-state wavefunction of the muonic deuterium atom at the
origin, and $\vert\psi_{nn}\rangle$ and $\vert\psi_d;s_d\rangle$ are,
respectively, the states of the $nn$ system and that of the
spin-polarized deuteron with projection $s_d$.  
 Here we have ignored the quartet channel of muonic deuterium and 
only coupled the muon and the deuteron spins to 1/2.
For capture into the
$^1S_0$ singlet $nn$ state with relative momentum $p$, the
differential doublet capture rate is given by
\begin{equation}
\label{eq:doublet_rate}
 \frac{\mathrm{d}\Gamma_D^{{}^1S_0}}{\mathrm{d}p} = \frac{1}{2\pi^3} p^2 E_\nu^2 \(1-\frac{E_\nu}{m_\mu+m_d}\)
 \overline{\vert T_{fi} \vert^2}\,,
\end{equation}
where the spin-averaged squared matrix element
$\overline{\vert T_{fi} \vert^2}$ can be obtained from
Eq.~\eqref{eq:tfi} by averaging over the spin projections $s_{\mu d}$
of the muonic deuterium atom and summing over neutrino helicities $h$,
which gives
\begin{align}
 \overline{\vert T_{fi} \vert^2} = \frac{1}{6} G_V^2\,\phi_{1S}^2 & (\bm{0})\,
 \vert\sqrt{2} \langle\psi_{nn}\vert J^1(\mathbf{q})-iJ^2(\mathbf{q}) \vert \psi_d; 1 \rangle\nonumber\\
 &\quad -\langle\psi_{nn}\vert J^0(\mathbf{q}) + J^3(\mathbf{q}) \vert \psi_d; 0 \rangle\vert^2.
\end{align}
The neutrino energy is $E_\nu = \frac{1}{2m_{\mu d}} \[ m_{\mu d}^2 -
  4\(m_n^2+p^2\) \]$, 
%is related to $p$ by 
%\begin{equation}
% E_\nu = \frac{1}{2m_{\mu d}} \[ m_{\mu d}^2 - 4\(m_n^2+p^2\) \],
%\end{equation}
where $m_{\mu d}$ and $m_n$ are the masses of the muonic deuterium
atom and the neutron, respectively.  The integrated capture rate
$\Gamma_D^{{}^1S_0}$ can be obtained by integrating
Eq.~\eqref{eq:doublet_rate} with respect to $p$ between the limits 0
and $p_\mathrm{max}=(m_{\mu d}^2/4-m_n^2)^{1/2}$.

% * Currents
\section{\bf Weak currents}
\label{sec:currents}
The expressions for the charge-changing nuclear electroweak currents,
$J^\alpha\equiv
V_\mathrm{1B}^\alpha+A_\mathrm{1B}^\alpha+V_\mathrm{2B}^\alpha+A_\mathrm{2B}^\alpha$,
have been derived in chiral effective field theory in
Refs.~\cite{Park:1995pn,Park:1998wq,Park:2002yp,Song:2008zf}. We take
into account operators that give non-vanishing contributions to
Eq.~\eqref{eq:reaction} up to $\mathcal{O}(Q^3)$ in the chiral
expansion.  The nuclear wavefunctions are also consistently calculated
up to the same order. In both, the current operators and the wavefunction, we count $Q/m$ as $\mathcal{O}(Q^2)$
\cite{Carlsson:2015vda,Menendez:2011qq}.  The Gamow-Teller operator,
\begin{equation}
\label{eq:gt}
 \mathbf{A}_\mathrm{1B}^\mathrm{GT}(\mathbf{q}) = -F_A(q^2) \sum_i e^{-i\mathbf{q}\cdot\mathbf{r}_i} \tau_i^-\bm{\sigma}_i\,,
\end{equation}
enters at $\mathcal{O}(Q^0)$. Here, $F_A$ is the axial form-factor
which is a function of the four-vector inner product
$q^2=m_\mu(m_\mu-2E_\nu)$. We use $F_A(q^2) = g_A(1+r_A^2q^2/6)$,
where $r_A$ is the axial radius of the nucleon.  This truncation is
consistent with the chiral order to which we work in this paper, and
with both dipole- and $z$-parameterizations of the axial form-factor
\cite{Meyer:2016oeg}. The pseudo-scalar operator~\cite{Park:2002yp},
\begin{equation}
\label{eq:ps}
A_\mathrm{1B}^0(\mathbf{q}) =
-g_A \sum_i e^{-i\mathbf{q}\cdot\mathbf{r}_i} \tau_i^-\frac{\bm{\sigma}_i\cdot\mathbf{\bar p}_i}{m}\,,
\end{equation}
where
$\mathbf{\bar
  p}_i=(\mathbf{p}_i+\mathbf{p^\prime}_i)/2=\mathbf{p}_i+\mathbf{q}/2$
is the average of the momenta of the nucleons before and after
coupling with the leptons, only appears at
$\mathcal{O}(Q^2)$. Additionally both Eqs.~\eqref{eq:gt} and
\eqref{eq:ps} also include an induced-pseudoscalar contribution
~\cite{Ando:2001es} that gives
$A_\mathrm{1B}^\alpha(\mathbf{q})\rightarrow
A_\mathrm{1B}^\alpha(\mathbf{q})+q^\alpha q_\beta
A_\mathrm{1B}^\beta(\mathbf{q})/(m_\pi^2-q^2)$. The one-body vector
operator appears at $\mathcal{O}(Q^2)$ and consists of the so-called
convection current and the weak-magnetism terms,
\begin{equation}
\label{eq:one_body_vector}
 \mathbf{V}_\mathrm{1B}(\mathbf{q}) = \sum_i e^{-i\mathbf{q}\cdot\mathbf{r}_i} \tau_i^- 
 \frac{1}{m}\(\mathbf{\bar p}+ i \frac{\mu_V}{2} \mathbf{q}\times\bm{\sigma}_i\)\,,
\end{equation}
where $\mu_V$ is the nucleon isovector magnetic moment, whose value is 
4.706. 
In Eqs.~\eqref{eq:ps} and \eqref{eq:one_body_vector}, and also in the 
two-body currents discussed below, we have 
used the zero four-momentum transfer values 
for the axial and electromagnetic form factors 
since their $q^2$ dependences are higher order in the EFT expansion. 

The axial two-body operators, which enter at $\mathcal{O}(Q^3)$, 
can be written as~\cite{Park:2002yp,Ando:2001es}
\begin{equation}
\label{eq:axail_2b_with_pion_pole}
 A_\mathrm{2B}^\alpha(\mathbf{q})=
\hat{A}_\mathrm{2B}^\alpha(\mathbf{q})+\frac{q^\alpha\[q_\beta \hat{A}_\mathrm{2B}^\beta(\mathbf{q})
+\hat A_\mathrm{2B}^\mathrm{PS}(\mathbf{q})\]}{m_\pi^2-q^2}~,
\end{equation}
where
\begin{align}
 \hat{A}_\mathrm{2B}^0(\mathbf{q}) &= -i\frac{g_A}{4f_\pi^2}\tau_\times^-
 \[\frac{\bm{\sigma}_1\cdot\mathbf{k}_1}{m_\pi^2-k_1^2} - \frac{\bm{\sigma}_2\cdot\mathbf{k}_2}{m_\pi^2-k_2^2}\]\nonumber\\
  & + \frac{2g_A}{mf_\pi^2}\(\hat c_2+\hat c_3-\frac{g_A^2}{8}\)\sum_i\tau_i^-\frac{\bm{\sigma}_i\cdot\mathbf{k}_ik_i^0}{m_\pi^2-k_i^2}\,,
\end{align}

\begin{align}
\label{eq:axail_2b_proper}
 \hat{\mathbf{A}}_\mathrm{2B}(\mathbf{q}) = &\frac{g_A}{2mf_\pi^2} \bigg\{\frac{\bm{\sigma}_2\cdot\mathbf{k}_2}{m_\pi^2-k_2^2}\bigg[
  \frac{i}{2}\tau_\times^-\mathbf{\bar p}_1 + 4\hat c_3 \tau_2^-\mathbf{k}_2\nonumber\\ 
 &+\(\hat c_4+\frac{1}{4}\)\tau_\times^-\bm{\sigma}_1\times\mathbf{k}_2 +\frac{\mu_V}{4}\tau_\times^-\bm{\sigma}_1\times\mathbf{q}\bigg] \nonumber\\
 &+2 \hat d_1 \sum_i \tau_i^-\bm{\sigma}_i + \hat d_2\tau_\times^-\bm{\sigma}_\times + (1\leftrightarrow 2)\bigg\}~,
\end{align}
and 
\begin{equation}
\label{eq:ps_2b}
\hat{A}_\mathrm{2B}^\mathrm{PS}(\mathbf{q})= \frac{4g_Am_\pi^2}{mf_\pi^2}\,\hat c_1 
 \[\tau_2^-\frac{\bm{\sigma}_2\cdot\mathbf{k}_2}{m_\pi^2-k_2^2}
 +(1\leftrightarrow2)\]~.
\end{equation}
The $\mu_V$ term in Eq.~\eqref{eq:axail_2b_proper} and the 
pion pole contribution given by the second term in Eq.~\eqref{eq:axail_2b_with_pion_pole} 
were ignored by Ref.~\cite{Park:2002yp} in their proton-proton fusion calculation 
but were included by Ref.~\cite{Ando:2001es} since they are non-negligible for the 
muon capture process. In these equations, $\mathbf{k}_i=\mathbf{p}^\prime_i-\mathbf{p}_i$,
$\tau_\times^- = (\tau_1\times\tau_2)^x - i (\tau_1\times\tau_2)^y$,
$\bm{\sigma}_\times =\bm{\sigma}_1\times\bm{\sigma}_2$ and $f_\pi$ is
the pion decay constant. The linear combination
$g_A\Lambda_b\,(\hat d_1 + 2 \hat d_2)=c_D$ is conventionally used to
combine the $\hat d_1$ and $\hat d_2$ terms, which are rendered
redundant by the Pauli principle~\cite{Park:2002yp}.  The LECs
$c_i$ in the pion-exchange current also appear in
$\pi$N and $NN$ interactions and in the long-range part of
the $NNN$ interaction, whereas $c_D$ (or $\hat d_R$) simultaneously
parameterizes both the strength of the short-range part of the
meson-exchange axial currents and that of the intermediate-range part
of the $NNN$ interaction.  The vector part of the two-body current is
given by the sum of the so-called seagull and pion-in-flight terms~\cite{Ando:2001es},
\begin{align}
\mathbf{V}_\mathrm{2B}(\mathbf{q})= -i\tau_\times^-& \frac{g_A^2}{4f_\pi^2}\bigg[
  \frac{\bm{\sigma}_1\bm{\sigma}_2\cdot\mathbf{k}_2}{m_\pi^2-k_2^2}-\frac{\bm{\sigma}_2\bm{\sigma}_1\cdot\mathbf{k}_1}{m_\pi^2-k_1^2}\nonumber\\
& +\frac{\bm{\sigma}_1\cdot\mathbf{k}_1}{m_\pi^2-k_1^2}\frac{\bm{\sigma}_2\cdot\mathbf{k}_2}{m_\pi^2-k_2^2}(\mathbf{k}_2-\mathbf{k}_1)\bigg]\,.
\end{align}

 The two-body vector charge operator, $V_\mathrm{2B}^0$, is suppressed by an additional factor of 
the chiral EFT expansion parameter. 

% ** Uncertainty quantification
\section{\bf Covariance analysis}
\label{sec:stat_analysis}
The covariance matrices provided in Ref.~\cite{Carlsson:2015vda} offer
a straightforward handle on the statistical uncertainties in the
integrated and differential muon-capture rate stemming from the
experimental uncertainties in the fit data. The Jacobians of of
$\Gamma_D^{{}^1S_0}$ with respect to relevant LECs were computed in a
simple finite difference scheme and derivatives could be reliably
extracted using splines. For the nuclear wavefunctions we do not allow
any variation in the axial coupling constant $g_A$. We start from
$g_A=1.276$~\cite{PhysRevLett.105.181803} which after renormalization
to account for the Goldberger-Treiman discrepancy is matched to the
empirically determined $\pi N$ coupling strength $g_{\pi N N}^2/4\pi =
13.7$~\cite{Baru}. This value for $g_A$ is slightly larger than the
most recently adopted Particle Data Group (PDG) value $g_A =
1.2723(23)$ but in fair agreement with the value $g_{A}=1.2749(9)$
employed by Hill {\it et al.} \cite{Hill:2017wgb}. 
%%%%%%%%%%%%%%%%%%%%%%%%%%%%%%%%%%%%%%%%%%%%%%%%%%%%%%%
\begin{figure}
  \label{fig:simpots}
  \includegraphics[width=\columnwidth]{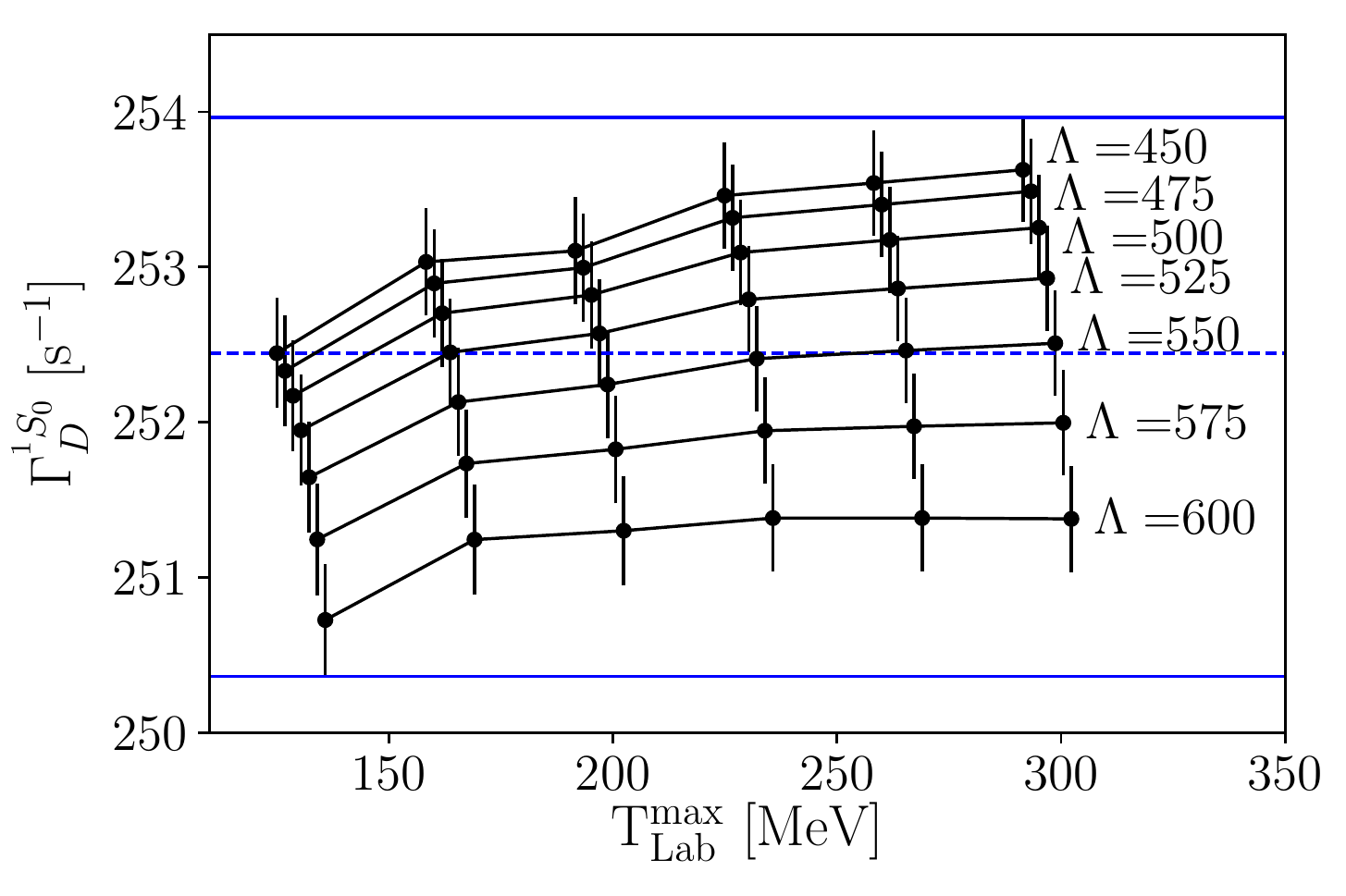}
  \caption{\label{fig:nnlo_sim} Distribution of central values for the
    muon-capture rate $\Gamma_D^{{}^1S_0}$ when using the family of 42
    chiral EFT potentials at order $Q^3$ from
    Ref.~\cite{Carlsson:2015vda}. The vertical bars indicate the
    respective statistical uncertainty propagated from the underlying
    uncertainties in the LECs. Calculations with identical regulator
    cutoffs $\Lambda$ but different truncations $T_{\rm Lab}^{\rm
      max}$ in the $NN$ scattering database are connected with a
    line to guide the eye. The weighted average of all
    calculations and conservative error limits are indicated with
    dashed and solid lines, respectively. The numerical values of the
    combined model error is given in Eq.~\eqref{eq:gamma_sim}.
  }
\end{figure}
%%%%%%%%%%%%%%%%%%%%%%%%%%%%%%%%%%%%%%%%%%%%%%%%%%%%%%%%%%%%

It is sufficient to use the first-order statistical methods described
in Ref.~\cite{Carlsson:2015vda}. From this we can establish that the
uncertainty in $\Gamma_D^{{}^1S_0}$ due to uncertainties in the
determination of the LECs at $Q^3$ from experimental data is very
small and certainly not of any primary concern. We find that the
typical size of the statistical uncertainties in $\Gamma^{{}^1S_0}_D$
is $0.5$ s$^{-1}$. The sensitivity to different truncations of the
$NN$ scattering database is of similar size, while variations of the
regulator cutoff is up to five times larger, see
Fig.~\ref{fig:nnlo_sim}

Based on the covariance analysis, variations of the regulator cutoff,
and the pool of fit data for extracting the LECs we obtain a
conservative estimate for the model uncertainty at order
$Q^3$ in chiral EFT. Using a weighted average of the results shown in
Fig.~\ref{fig:nnlo_sim}, we find
\begin{equation}
  \label{eq:gamma_sim}
  \Gamma^{{}^1S_0}_D = 252.4^{+1.5}_{-2.1}~s^{-1}.
\end{equation}

% ** Observables and Correlations
\section{\bf Correlation with the astrophysical proton-proton $S$-factor}
\label{sec:corr_analysis}

\begin{figure}
  \includegraphics[width=\columnwidth]{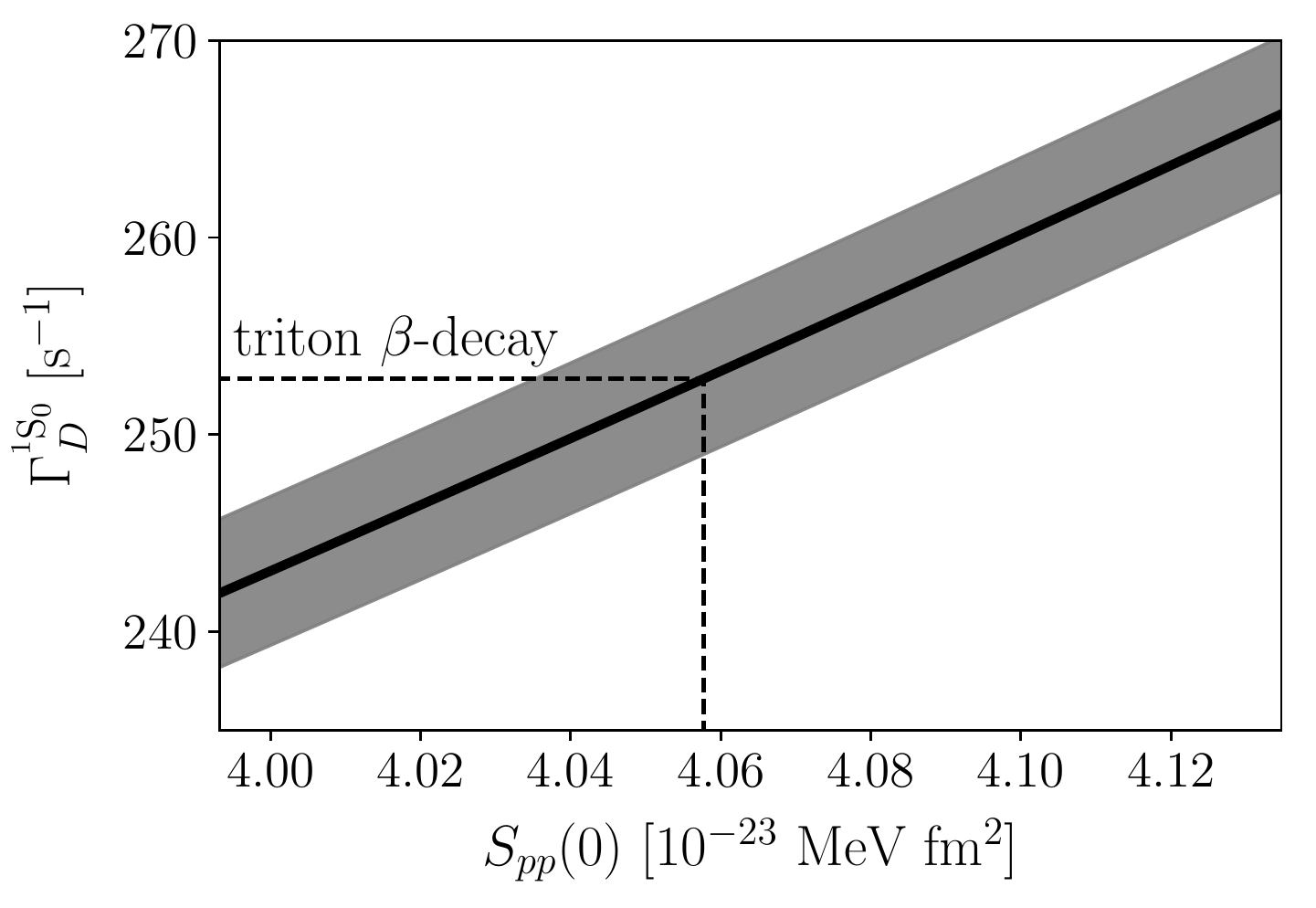}
  \caption{\label{fig:cor} The muon-capture rate $\Gamma_D^{{}^1S_0}$ as a
    function of the $pp$ fusion $S_{pp}$-factor at zero energy
    parameterized by the axial current LEC $c_D\in[-3.6,+3.6]$ at
    order $Q^3$ using the Roy-Steiner based interaction. The grey band
    indicates the uncertainty in the muon-capture rate due to the
    uncertainty in the axial radius of the nucleon $r_A^2=0.46(22)$
    fm$^2$. The dashed lines indicates the values for $S_{pp}(0)$ and
    $\Gamma_D$ when the experimental value for the triton
    $\beta$-decay half-life is used to determine the LEC $c_D$.}
\end{figure}

Using the chiral interactions at orders $Q^3$ with the $\pi N$ LECs,
$c_{1,2,3,4} = (-0.74,1.81,-3.61,2.44)~\mathrm{GeV}^{-1}$, determined
in a Roy-Steiner analysis~\cite{Siemens:2017jb}, we can analyze the
relation between different observables via a variation of the
short-distance LEC $c_D$. For example, in Fig.~\ref{fig:cor} we trace
out the correlation between the proton-proton ($pp$) $S_{pp}$-factor
at zero energy and the muon-capture rate
$\Gamma^{{}^1S_0}_D$.  Different points on the black line in this
figure only differ in the values of the LEC $c_D$ and $c_E$ that
reproduce the binding energies and radii of $^{3}$H and $^{3}$He while
the two-body interaction remain unchanged.

 As in Ref~\cite{0000000002995761}, we can also use the triton binding
 energy and $\beta$-decay half-life, corresponding to a reduced matrix
 element of the $J=1$ electric multipole of the axial-vector current
 $|\langle ^3{\rm He} || E_{1}^{A}|| ^{3}H \rangle| = 0.6848\pm0.0011$, to fix
 $c_D$ and $c_E$, and thus also the muon-capture rate and the $pp$
 fusion $S$-factor, see Fig.~\ref{fig:cor} (dashed lines).  With
 $c_D=-0.39$ and $c_E=-0.44$, we find that $S_{pp}(0) = 4.058 \times
 10^{-23}$ MeV fm$^2$ , which is consistent with our previously
 published result~\cite{Acharya:2016kfl}, and a muon-capture rate
\begin{align}
  \label{eq:gamma_rs}
  \Gamma^{{}^1S_0}_{D} &= 252.8 \pm 4.6 \pm 3.9 ~s^{-1}~.
\end{align}

These results are not sensitive to variations of the tritium $\beta$-decay
matrix element within the range of uncertainty quoted above.
The first uncertainty in the above expression for $\Gamma^{{}^1S_0}_{D}$ estimates the effect of
truncating the chiral EFT expansion at order $Q^3$. The second
uncertainty indicates the sensitivity to variations of the axial
radius within the error budget $r_A^2=0.46(22)$
fm$^2$~\cite{Meyer:2016oeg}. The truncation error is extracted by
following the method discussed in Ref.~\cite{Furnstahl:2015rha}. In
brief, we calculate the capture rate at the lower orders $Q^0$ and
$Q^2$, in the currents as well as the wavefunctions, and express the
results as an expansion of the form $\Gamma^{{}^1S_0}_{D} =
\Gamma_{\rm LO}^{{}^1S_0} \sum^{3}_{n=0}c_n (p/\Lambda_b)^n$, where we
assume that the breakdown scale of theory is $\Lambda_b = 500$~MeV and
the inherent momentum $p$ of the problem is provided by the soft scale
of chiral EFT, i.e. $p=m_{\pi}$. Note that the maximum of the
momentum-differential doublet-capture rate in
Eq.~\ref{eq:doublet_rate} occurs at a momentum scale $p\sim 25$
MeV. We obtain an estimate for the EFT truncation error by calculating
%\begin{equation}
$(p/\Lambda_b)^4 {\rm \, max}(|c_0|,|c_2|,|c_3|)$~.
%\end{equation}
The order-by-order capture rates with a $c_D$ that reproduces the
comparative inverse $\beta$-decay half-life of triton are $(186.3,247.3,252.8)$
$s^{-1}$ at orders $(Q^0,Q^2,Q^3)$, respectively.
We find that the uncertainty estimate resulting from an
analysis of the EFT truncation is comparable to the error induced by
the imprecise value of the axial radius. In turn, both of these errors
are twice as large as the uncertainty related to the cutoff
variation of the chiral potential and truncations in the pool of fit
data.

%%%%%%%%%%%%%%%%%%%%%%%%%%%%%%%%%%%%%%%%%%%%%%%%%%%%%%%%%%%%
%%%%%%%%%%%%%%%%%%%%%%%%%%%%%%%%%%%%%%%%%%%%%%%%%%%%%%%%%%%%%

% * Conclusion
\section{\bf Conclusion}
\label{sec:sconclusion}
We have analyzed uncertainties in calculations for the muon-capture
rate using two classes of interactions: (i) order $Q^3$ interactions
constructed as described above and in Ref.~\cite{Carlsson:2015vda},
and (ii) a set of interactions at order $Q^0,Q^2,Q^3$, whose $\pi N$
couplings $c_1$, $c_3$ and $c_4$ were taken from
Ref.~\cite{Siemens:2017jb}.

The analysis carried out in Ref.~\cite{Hoferichter:2015hva,Siemens:2017jb} has
reduced the uncertainties in the $\pi N$ LECs significantly. This
leaves $c_D$ as the only undetermined LEC in the weak axial two-body
current. We demonstrated that this leads to linear correlations
between electroweak observables in the two-nucleon sector that involve
phaseshift equivalent $NN$ interactions.

We focused on the singlet $S$-wave $nn$ channel, which is the only
channel sensitive to the weak axial two-body contact current. Our results for
muon-capture and the associated uncertainties are shown in
Eqs.~\eqref{eq:gamma_sim} and \eqref{eq:gamma_rs}. These uncertainty
estimates are rooted in the description of the strong-interaction
part of the calculation. We also emphasize the importance of the
additional $\sim 1.5\%$ uncertainty due to the uncertainty
in the nucleon axial radius. We note that the central value 
we obtain is in excellent agreement with a prior chiral EFT calculation~\cite{Marcucci:2011jm} 
even though our error estimate is larger because we perform a more rigourous treatment of 
uncertainties.

Using the result for muon-capture into the single $S$-wave from
Eq.~\eqref{eq:gamma_rs} and the results from
Ref.~~\cite{Marcucci:2011jm} for muon-capture into higher partial waves, we
can obtain an estimate of $397.8~$s$^{-1}$ for the total capture rate, 
$\Gamma_D$. We expect that higher accuracy can
be obtained for capture into higher partial waves since these are less
sensitive to the axial two-body current. However, we refrain from
giving a total uncertainty for this capture rate.

We have also studied the correlation of the capture rate with other
$NN$ observables. In agreement with previous
work~\cite{Marcucci:2014uoa}, we find that the capture rate depends
only weakly on the $nn$ scattering length $a_{nn}$ provided
that it is negative. However, the capture rate would be significantly
smaller if $a_{nn}$ was positive due to the existence of a shallow
dineutron.

In the future, we will carry out a complete uncertainty analysis for
$pp$ fusion and muon capture on the deuteron, including the effect of
higher partial waves. This analysis will provide a full picture on
the uncertainties and correlations of electroweak processes in the
$NN$ sector. We emphasize that the axial radius $r_A$ is a
significant source of uncertainty in our analysis.
Future improvements in
experimental precision and lattice QCD results~\cite{Chang:2018uxx}
will lead to important insights into how the nuclear Hamiltonian
correlates various electroweak observables.

\section{\bf Acknowledgments}
This work has been supported by the National Science Foundation under
Grant No. PHY-1555030, by the Office of Nuclear Physics,
U.S.~Department of Energy under Contract No. DE-AC05-00OR22725, and 
by the Deutsche Forschungsgemeinschaft through The Low-Energy
Frontier of the Standard Model (SFB 1044) CRC and through the PRISMA
Cluster of Excellence. This project has received funding from the
European Research Council (ERC) under the European Union's Horizon
2020 research and innovation programme (grant agreement No 758027) and
the Swedish Research Council under Grant No. 2015-00225 and Marie
Sklodowska Curie Actions, Cofund, Project INCA 600398. 
The computations were performed on resources provided by the Swedish National Infrastructure for Computing at NSC (Project SNIC 2018/3-346).

% * Bibliography
%====================  Bibliography ==============================

\end{document}